\documentclass[aip,jcp,graphicx,sd,reprint]{revtex4-1}
\usepackage{graphicx}
\usepackage{dcolumn}
\usepackage{bm}
\begin{document}

\title{Solid-state diffusion in amorphous zirconolite}
\author{C. Yang}
\address{School of Physics and Astronomy, Queen Mary University of London, Mile End Road, London, E1 4NS, UK}
\author{E. Zarkadoula}
\address{School of Physics and Astronomy, Queen Mary University of London, Mile End Road, London, E1 4NS, UK}
\address{Materials Science and Technology Division, Oak Ridge National Laboratory, Oak Ridge, TN 37831-6138, USA}
\author{M. T. Dove}
\address{School of Physics and Astronomy, Queen Mary University of London, Mile End Road, London, E1 4NS, UK}
\author{I. T. Todorov}
\address{STFC Daresbury Laboratory, Warrington WA4 1EP, UK}
\author{T. Geisler}
\address{Steinmann-Institut f\"{u}r Geologie, Mineralogie und Pal\"{a}ontologie, University of Bonn, D-53115 Bonn, Germany}
\author{V. V. Brazhkin}
\address{ Institute for High Pressure Physics, RAS, 142190, Moscow, Russia}
\author{K. Trachenko}
\address{School of Physics and Astronomy, Queen Mary University of London, Mile End Road, London, E1 4NS, UK}

\begin{abstract}
We discuss how structural disorder and amorphization affects solid-state diffusion, and consider zirconolite as a currently important case study. By performing extensive molecular dynamics simulations, we disentangle the effects of amorphization and density, and show that a profound increase of solid-state diffusion takes place as a result of amorphization. Importantly, this can take place at the same density as in the crystal, representing an interesting general insight regarding solid-state diffusion. We find that decreasing the density in the amorphous system increases pre-factors of diffusion constants, but not decreasing the activation energy. We also find that atomic species in zirconolite are affected differently by amorphization and density change. Our microscopic insights are relevant for understanding how solid-state diffusion changes due to disorder and for building predictive models of operation of materials to be used to encapsulate nuclear waste.
\end{abstract}


\maketitle

\section{Introduction}

A variety of radiation sources are created and used in science and technology. This includes an important area of energy generation in nuclear power stations, where kinetic energy of fission products is converted into heat and electricity. In this and other applications, the energy of emitted particles often has a two-fold effect. On one hand, this energy is converted into useful energy, by heating the material around the particle tracks. On the other hand, this energy damages the material and degrades its important properties, including mechanical, thermal, transport and so on, to the point that a material might lose its functional purpose. A problem in fission and future fusion reactors \cite{stoneham}, this issue is particularly acute in the process of safe immobilization of nuclear waste, and constitutes one of the pressing issues that modern society faces \cite{web-rev,web-rev0}. The issue is closely related to public acceptance of nuclear industry and therefore to the future of nuclear power. Regardless of the future of nuclear industry, the amount of accumulated nuclear waste is very large and is steadily growing while no acceptable solution of its safe storage exists.

Crystalline ceramics have been proposed to immobilize highly radioactive nuclear waste \cite{web-rev,web-rev0,synroc1,synroc2,report}. The main requirement for the immobilization matrix, the waste form, is to prevent the radioactive isotopes from leaking and polluting the environment. This is perceived to be a very challenging requirement, given the high radioactivity of the nuclear waste and the long radioactivity time spans that extend from several thousand years for fission products to several million years for actinides. As recently observed \cite{report}, nuclear waste encapsulation involves ``design problem the likes of which humanity had never before attempted, because it involved a time scale that required predictions of material and system behavior tens and hundreds of thousands of years into the future. Some perspective on the uniqueness of this temporal projection comes from the realization that the most ancient monuments of past engineering achievement, such as Stonehenge and the Pyramids, are barely five thousand years old.'' With no direct testing possible over long periods of time, the decision of using a particular wasteform needs to be informed by indirect simulated experiments as well as detailed theoretical understanding of how irradiation affects the ability of the waste form to remain an effective immobilization barrier.

Waste form alteration, the main concern for the safe encapsulation of nuclear waste, is a complex phenomenon involving diffusion, leaching and dissolution \cite{str2,thor-rev}. Understanding the effects of irradiation on waste form durability is challenging because the process is complex and includes many mechanisms at work \cite{web-rev,web-rev0,thor-rev}. Generally, wasteform alteration as a result of irradiation can include chemical changes at the surface, reactions with water and other environmental agents, increase of solid-state diffusion of atoms in the bulk due to radiation damage, increased defect mobility, defects interaction and so on. The above processes can be system specific, yet have a common underlying mechanism, thermal diffusion. It is therefore important to understand diffusion in waste forms and its changes due to radiation damage.

An important effect of irradiation on the waste form is the remarkable increase of diffusion as a result of radiation-induced amorphization \cite{gei1,gei2}. Seen in easily amorphizable materials, this effect has been thought to be either absent or reduced in materials resistant to amorphization by radiation damage \cite{resist}. However, even most resistant materials such as ZrO$_2$ still show considerable damage in the form of a large number of well-separated point defects \cite{zro2}.

An interesting possibility is having a waste form which is amorphized by radiation damage yet still shows low levels of alteration and continues to be an effective immobilization barrier. Zirconolite, ZrCaTi$_2$O$_7$, is one of the phases in SYNROC mixture of different ceramics each tailored to immobilize different ions present in the highly-radioactive nuclear waste \cite{synroc1,synroc2}, and is currently the preferred waste form by the UK National Decommissioning Agency for immobilization of actinides. In more recent experiments \cite{str1,str2,str3}, zirconolite has been rendered X-ray amorphous by alpha-decay processes of Pu and, surprisingly and contrary to other materials, did not reveal significant chemical and physical alterations, witnessed by the absence of phase changes and microcracks even at fairly large volume increases. Furthermore, zirconolite maintained strong elastic response and overall chemical durability \cite{str1,str2,poml}, in contrast to other materials studied. At the same time, the aqueous durability of zirconolite is strongly affected by radiation damage \cite{poml}, with the evidence supporting diffusion-controlled ion exchange as the main mechanism of alteration of radiation-damaged zirconolite. These results call for further detailed investigation of the mechanisms involved in the alteration of this wasteform and diffusion mechanism in particular.

According to current understanding, the increase of diffusion due to radiation-damaged system is due to the associated density decrease. Indeed, apart from rare examples such as Si, radiation-induced structural changes and amorphization are accompanied by density decrease. The effect of density on the activation energy for diffusion $U$ (the energy needed by an atom to jump from its surrounding cage to the neighbouring quasi-equilibrium location) has been well understood since the early work of Frenkel \cite{frenkel}. When interatomic separations are at their equilibrium values in a solid, $U$ is too high for the jump event to take place during any feasible time. However, if the cage can increase its size (for example, due to thermal fluctuations) and open up a low-energy local diffusion pathway, the jump can proceed much faster. If $\Delta r$ is the increase of the cage size required for the jump to take place, $U$ is equal to the work required to expand the change elastically, and is

\begin{equation}
U=8\pi G r\Delta r^2
\end{equation}

\noindent where $G$ is shear modulus and $r$ is the cage radius \cite{frenkel}. Note that the elastic energy to expand the sphere of radius $r$ by amount $\Delta r$ depends on shear modulus $G$ only because no compression takes place at any point. Instead, the system expands by the amount equal to the increase of the sphere volume \cite{frenkel}, resulting in a pure shear deformation. Indeed, the strain components $u$ from an expanding sphere (noting that $u\rightarrow$0 as $r\rightarrow\infty$) are $u_{rr}=-2b/r^3$, $u_{\theta\theta}=u_{\phi\phi}=b/r^3$ \cite{landau}, giving pure shear $u_{ii}=0$.

From a theoretical standpoint, density is considered to be the main factor governing $G$ and hence $U$ \cite{frenkel}. Indeed, when density decreases as a result of radiation-induced structural changes, the interatomic interaction strength decreases, reducing $G$. Further, $\Delta r$ decreases due to the increase of the cage volume. Therefore, if $G$ is constant at constant density, Eq. (1) makes two predictions. First, $U$ reduces due to density decrease, the widely anticipated result corroborated by more recent work on diffusion processes in glasses and viscous liquids \cite{dyre}. Second, $U$ does not change at constant density.

In contrast to the density effect, the consequences of amorphization at {\it constant density} for diffusion are not understood. Indeed, if a structural change (e.g. amorphization or large accumulation of point defects and their clusters as in ZrO$_2$ \cite{zro2}) takes place at constant density, the volume of the atomic cage around the diffusing atom does not change on average. However, the wide distribution of interatomic angles in the disordered structure leads to the appearance of both faster and slower local diffusion pathways (see Fig. 1), even if this structure is of the same density as the parent crystal, with the net effect of increasing the diffusion. This is an agreement with experimental results reporting the decrease of $G$ as a result of amorphization at the same density \cite{brazhkin}.

Hard to estimate theoretically, the combined effect of local fast and slow diffusion pathways at the same density is important to understand from the waste form perspective \cite{thor-rev} as well as from the general point of view of properties of disordered state. Indeed, since the early work \cite{ziman}, the extent to which structural disorder affects system properties remains widely debated. For example, recent work aiming at resolving the long-standing debate about the origin of the Boson peak in the energy spectrum of glasses, has found that, contrary to earlier expectations, the difference of vibrational spectra and other important properties between the amorphous system and its parent crystal disappears once the densities of both systems are taken to be the same \cite{boson}. Generally, recently accumulated evidence suggests that disorder leaves some important properties of the system surprisingly unaffected, but modifies other properties substantially \cite{review}.

In this paper, we address an important element of waste form alteration, the change of thermal diffusion due to radiation damage in the waste form. We use molecular dynamics (MD) simulations to study the change of diffusion of different atomic species in zirconolite as a result of structural disorder. Importantly, MD simulations enable us to disentangle the effects of amorphization and density increase on diffusion and discuss these effects separately. Such a separation is very hard to achieve in experimental radiation-damaged samples. We find that a profound increase of solid-state diffusion takes place as a result of amorphization. Importantly, this can take place at the same density as in the crystal, representing an interesting general insight regarding solid-state diffusion. We find that increasing the volume in the amorphous system increases pre-factors of diffusion constants. We also find that atomic species in zirconolite are affected differently by amorphization and density change. Our findings are relevant for both understanding solid-state diffusion in the presence of disorder and for building predictive models of operation of nuclear waste forms.

\section{Results}

There are several ways in which structural disorder can be introduced. Although direct simulation of collision cascades gives detailed information about the nature of the damage, producing completely disordered structures by multiple cascade overlaps is not practical, especially for realistic high-energy events and large system sizes \cite{iron}. We prepared the amorphous structures by first melting the system at 5000 K, equilibrating the high-temperature liquid for 100 ps and subsequently quenched the liquid slowly to room temperature, 300 K. We note that amorphization by quenching the liquid can be different from radiation-induced amorphization in several respects \cite{gei1,gei2,henry1,henry2}, however our main motivation is to address a fundamental question of how diffusion is affected by topological disorder in general.

We use DL\_POLY MD simulation package \cite{dlpoly} and the system of 1056 atoms with the recent interatomic potential fitted to zirconolite properties \cite{poten}. We have simulated and relaxed three different zirconolite structures: crystalline, high-density amorphous with density equal to the crystal and low-density amorphous zirconolite with 5\% decreased density as in the experimental samples damaged by the radiation damage \cite{web-rev0,web0}. 10\% of U atoms were introduced as a substitution for Zr atom, representing a typical waste load in waste forms. The interatomic potential for the U--O interaction was taken from Ref. \cite{catlow}.

Experiments on alteration of damaged waste forms are conducted at high temperature in order to observe alteration and diffusion during laboratory time scale \cite{gei1,gei2,str1,str2,str3,thor-rev,web0}. Similarly, we performed several MD simulations at temperature high enough to observe diffusion. Diffusion was observed as the linear time dependence of the mean square displacement, $\langle r^2\rangle=6Dt$, where $D$ is the diffusion coefficient.

In Figure 2, we show representative $\langle r^2\rangle$ calculated for crystalline and amorphous zirconolite. We note that at short time, $\langle r^2\rangle$ crosses over from the oscillatory to the diffusive regime, acquiring the linear time dependence characteristic of diffusion (see Figures 2--3). We perform our subsequent analysis on the basis of the linear $\langle r^2\rangle\propto t$ diffusive regime at long times. We limit the analysis to 5 ns in time since at longer times at high temperature we observe the signatures of recrystallization, witnessed by the appearance of peaks in Zr--Zr and Ti--Ti sublattices beyond the medium-range order. For the following analysis, we will use two equations for temperature dependence of diffusion coefficient $D$ and hopping time $\tau$, the average time between two consecutive atomic jumps at one point in space \cite{frenkel}:

\begin{equation}
D=D_0\exp\left(-\frac{U}{{\rm k_B}T}\right)
\label{dif}
\end{equation}

\begin{equation}
\tau=\tau_{\rm D}\exp\left(\frac{U}{{\rm k_B}T}\right)
\label{tau}
\end{equation}
\noindent where $U$ is the activation barrier for the diffusion event, $\tau_{\rm D}$ is the shortest (Debye) vibration period of about 0.1 ps and pre-factor $D_0$ is the high-temperature limit of the diffusion coefficient when $\tau\rightarrow\tau_{\rm D}$.

Apart from O atoms, we observe no diffusion in the crystalline systems on the time scale of our simulations. On the other hand, we observe the diffusion of all atoms in the amorphous systems at the same temperature (see Figures 2--3). Importantly, this includes the diffusion in amorphous structures of the same density as the parent crystal. We therefore find that structural disorder at the same density increases the diffusion constant profoundly, an unexpected finding since the early work \cite{frenkel}, it was density which was believed to govern the activation energy barrier for diffusion \cite{dyre}. This represents our first main result from this work.

We next calculate main parameters of diffusion, $U$ and $D_0$, in amorphous systems and their change due to different density. We calculate $U$ by fitting the data in Figure 4 to Eq. (\ref{dif}) as $\ln D=\ln D_0-\frac{U}{T}$, and show the results in Table 1. The calculated values of $U$ in amorphous zirconolite represent our next quantitative result enabling future prediction of how the waste form will operate during long time scales, as discussed below in more detail.

In Table 1, we observe no appreciable differences of $U$ between amorphous structures of different densities within the error due to scatter. The scatter is related to large fluctuations in the system at very high temperature we simulated, and is larger for the less numerous U atoms. We therefore find that the considered moderate density decrease, corresponding to the experimental swelling of radiation-damaged zirconolite \cite{web-rev0,web0}, does not have a significant effect on $U$.

Since we do not observe diffusion in the crystalline zirconolite apart from O atoms, we are unable to calculate $U$ in the crystal and compare it to the amorphous system. However, we can estimate the lower bound of $U$ in the crystal, $U_l$, using Eq. (\ref{tau}): $U_l=T_h\ln\frac{\tau_l}{\tau_{\rm D}}$, where $\tau_l$ is the longest simulation time and $T_h$ is the highest temperature simulated in the crystal. Taking $T_l=2100$ K and $\tau_l=130$ ns, we find $U_l$ of about 3 eV. $U_l$ is therefore of the same order of magnitude as typical $U$ in crystals (typical $U$ in crystals can be larger by up to about a factor of 2 for different atoms).

A discernible trend in Figure 4 is the increase of pre-factor $D_0$ at smaller density. We estimate this increase using two methods. First, we directly calculate the increase of $D_0$ by fitting the MD data to $\ln D=\ln D_0-\frac{U}{T}$. This gives $\frac{D_{02}}{D_{01}}$ in the range $2-3$ for different atoms, where subscripts 1 and 2 refer to amorphous systems at larger (crystalline) and smaller density, respectively (see Table 1). Second, the range of $\frac{D_{02}}{D_{01}}$ can be estimated by subtracting $\ln D_1=\ln D_{01}-\frac{U_1}{T}$ and $\ln D_2=\ln D_{02}-\frac{U_2}{T}$. Using our earlier result that $U$ are the same within the error, we find $\ln\frac{D_{02}}{D_{01}}=\ln\frac{D_2}{D_1}$. Then, the range of $\ln\frac{D_{01}}{D_{02}}$ by calculating $\ln\frac{D_2}{D_1}$ at low and high temperature in Figure 4, giving $\frac{D_{02}}{D_{01}}$ in the range similar to the first method, as follows from Table 1. We therefore find that the increase of diffusion pre-factors with system's volume is appreciable.

As mentioned above, O atoms stand out from the rest of atomic species in that their diffusion is seen in the crystalline zirconolite, representing an interesting heterogeneity of atomic species in terms of diffusion. Weakly bound as compared to other atomic species, O atoms do not show discernible decrease of $U$ as a result of amorphization and density decrease (with the error quantified in Table 1 and present in Figures 2, 4). However, we observe the increase of $D_0$ as a result of amorphization by the factor of about 3. Contrary to other atoms, however, this factor is not sensitive to density decrease (see Figure 4 and Table 1).

\section{Discussion and summary}

There are several important insights from this work. First, we find that structural disorder and amorphization in particular, introduced to the system at the {\it same} density as the parent crystal, can result in a profound increase of solid-state diffusion. This is an interesting general insight in view that both earlier and current theories emphasize density as the main factor controlling the diffusion \cite{frenkel,dyre}. Our finding implies that the net effect of the appearance of slower and faster local diffusion pathways in the disordered structure (see Figure 1) is diffusion increase.

According to our result, faster local diffusion pathways dominate over slower ones in the same-density disordered structure. This is analogous to the well-known case of introducing equal amounts of harder and softer inclusions in an elastic matrix, with the result that the overall elastic response is mostly governed by the softer phase (in a simple example, this is illustrated by the inverse sum rule for bulk or shear moduli). The analogy is further relevant here because lower-density faster diffusion pathways and higher-density slower pathways can be approximately viewed as local regions that are softer and harder, respectively (see Figure 1). Then, the net effect is elastic softening of the system and hence smaller $G$, in agreement with experimental results \cite{brazhkin}. This implies smaller $U$ according to Eq. (1) and hence larger $D$ according to Eq. (2), consistent with our current findings.

We note here that contrary to the role of density which is well understood and quantified, the effect of disorder at constant-density is not amenable to a simple treatment or a model, but is important in a wider context of understanding the essential differences between crystalline and amorphous systems. For example, it has been long thought that amorphous systems are notably different in terms of their vibrational spectra, yet recent evidence has found that this is not the case if the amorphous system is at the same density \cite{boson}. This is consistent with a wider picture emerging that many important thermodynamic properties of the system are insensitive to disorder due to the similarity of their spectra whereas other properties, namely the transport properties such as thermal conductivity, are strongly affected \cite{review}.

Second, the increase of diffusion due to amorphization is important for the operation of waste forms. For example, under the typical waste load zirconolite will become amorphous from irradiation after about 1000 years. Being a very small fraction of time of operation of the waste form (100,000-1,000,000 years), this implies that solid-state diffusion will take place almost entirely in the amorphous state. This will take place with the associated increased diffusion constants that we have found in this work. Our results, and the calculated values of $U$ in particular, can therefore be used to predict the solid-state diffusion in the waste form during the most important period of its operation.

Third, we have found specifically that the combined effect of amorphization and volume increase can affect both the activation energy and diffusion pre-factors, but that different atomic species can be differently affected. For example, the diffusion of most numerous and mobile O atoms is not affected by amorphization to the same extent as in other atomic species.

In summary, by using MD simulations we have disentangled the effects of amorphization and density on solid-state diffusion, and showed that contrary to existing theories, a profound increase of diffusion takes place as a result of amorphization at the same density. We have found that decreasing the density in the amorphous system increases the pre-factors of diffusion constants. We have also found that atomic species in zirconolite are affected differently by amorphization and density change. Our microscopic insights are relevant for understanding how solid-state diffusion changes due to disorder and for constructing predictive physics-based models aimed at predicting the performance of waste forms over long time scales.

This research utilised Queen Mary's MidPlus computational facilities, supported by QMUL Research-IT and funded by EPSRC grant EP/K000128/1.We are grateful to E. Maddrell for discussions and to CSC for support.

\clearpage

\begin{table*}
\caption{\label{tab:table1} Activation energy $U$ (eV) of atomic species in the crystalline and amorphous zirconolite at two different densities. Pre-factors $D_0$ (m$^2$/s) are denoted as $D_{01}$ (amorphous system at larger crystalline density), $D_{02}$ (amorphous system at smaller density) and $D_{\rm crystal}$ (pre-factor of difusion of O atoms in the crystal). The change of diffusion pre-factors is evaluated using two methods discussed in the text: calculating $D_0$ from direct fitting to calculated $D$ (a) and estimating the range of $D_0$ at different temperature (b). Subscripts 1 and 2 refer to amorphous systems at larger (crystalline) and smaller density, respectively.}
\begin{tabular}{p{3cm} p{2cm} p{2cm} p{2cm} p{2cm} p{2cm}}
\hline
&Ca&Zr&Ti&U&O \\
\hline
$U$ (amorphous, same density as crystal) & $2.80\pm0.53$ & $3.05\pm0.73$ & $3.79\pm0.63$ & $3.20\pm0.86$ & $1.61\pm0.06$ \\
$U$ (amorphous, smaller density) & $2.80\pm0.20$ & $3.05\pm0.71$ & $3.79\pm0.55$ & $3.20\pm0.54$ & $1.61\pm0.15$ \\
$U$ (crystal) & & & & & $1.61\pm0.07$ \\
(a)\\
$\ln\frac{D_{02}}{D_{01}}$ & $0.91\pm0.09$ &$0.66\pm0.15$ & $0.76\pm0.13$ & $1.01\pm0.18$ & $0.0\pm0.2$ £\\
$\ln\frac{D_{01}}{D_{\rm crystal}}$ & & & & & $1.04\pm0.1$ \\
(b) \\
$\ln\frac{D_{02}}{D_{01}}$ & $0.89-0.92$ &$0.62-0.71$ & $0.54-1.00$ & $0.79-1.27$ & \\
$\ln\frac{D_{01}}{D_{\rm crystal}}$ & & & & & $0.94-1.11$ \\
\hline
\end{tabular}
\end{table*}

\begin{figure}
{\includegraphics[width=8cm]{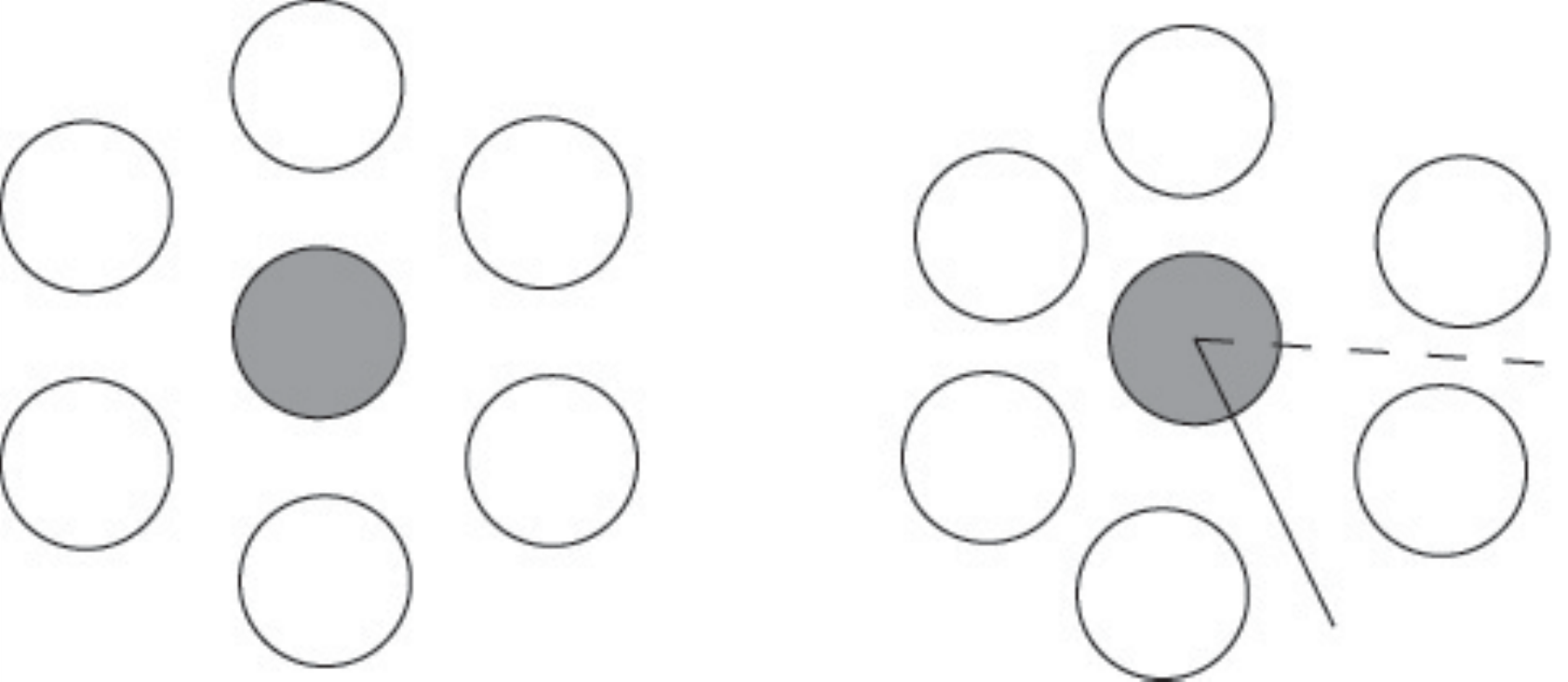}}
\caption{Schematic picture showing equivalent interatomic distances and local diffusion pathways in the crystalline structure (left). The disordered structure with the same density as the parent crystal has a wide distribution of interatomic distances and angles and gives rise to both faster (solid line) and slower (dashed line) local diffusion pathways.}
\label{1}
\end{figure}

\begin{figure}
{\includegraphics[width=8cm]{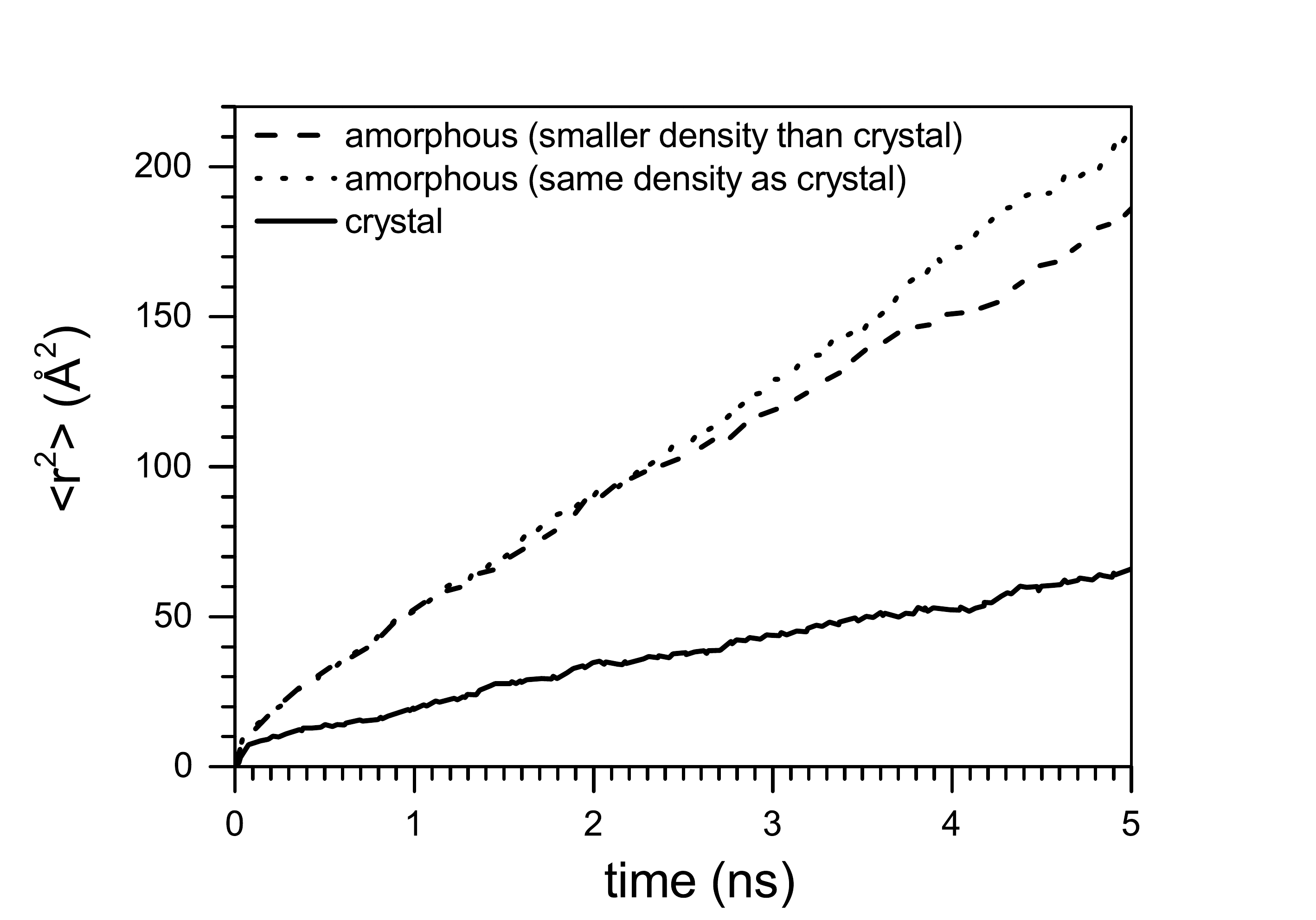}}
\caption{Mean-squared displacement of O atoms (averaged over all atoms) in crystalline and amorphous zirconolites of two different densities at 2000 K.}
\label{2}
\end{figure}

\begin{figure}
{\includegraphics[width=8.5cm, trim=1cm 0 0 0]{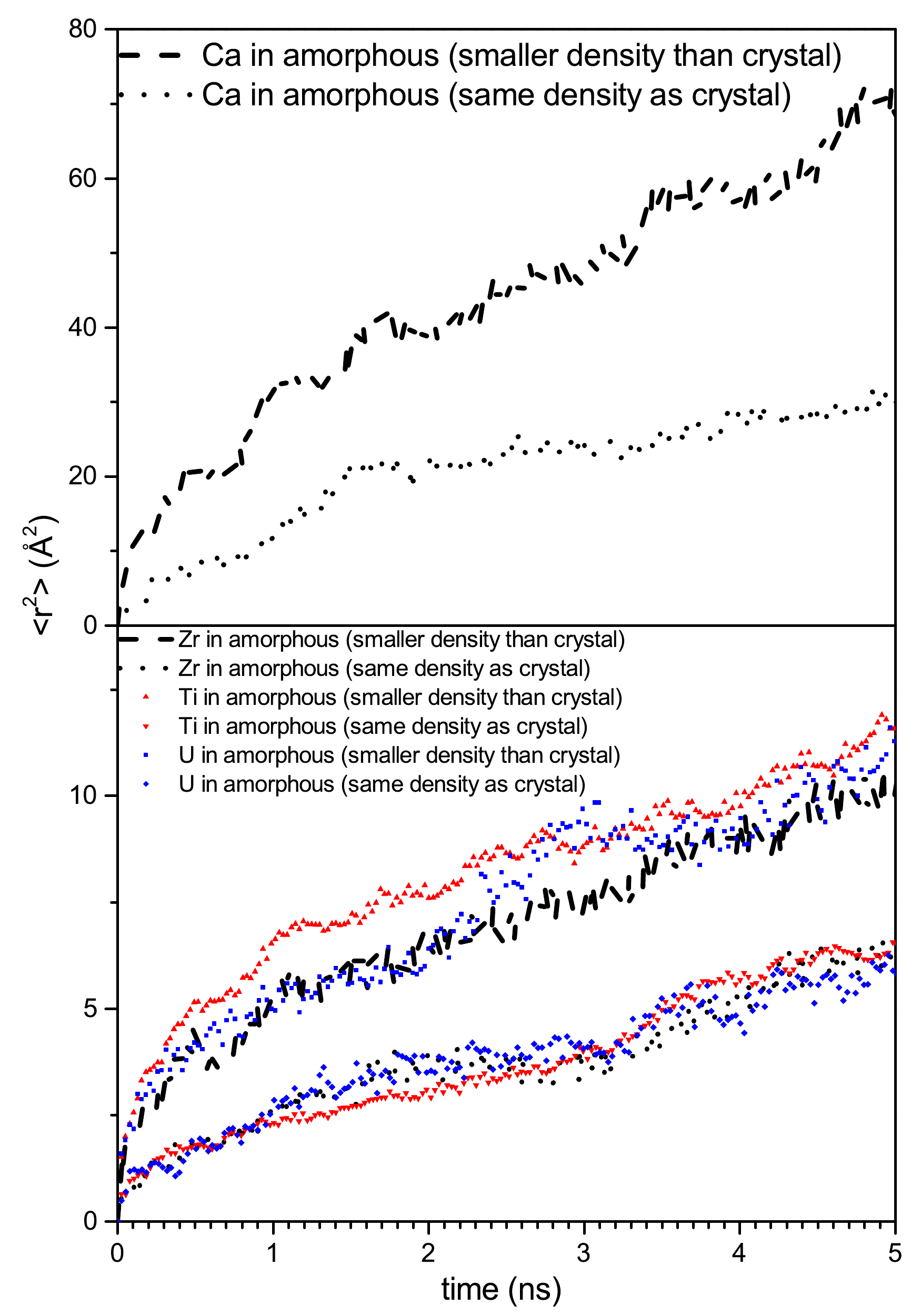}}
\caption{Mean-squared displacement of Zr, Ti, Ca and U atoms (averaged over all atoms) in amorphous zirconolites of two different densities at 2000 K.}
\label{3}
\end{figure}

\begin{figure}
{\includegraphics[width=8cm]{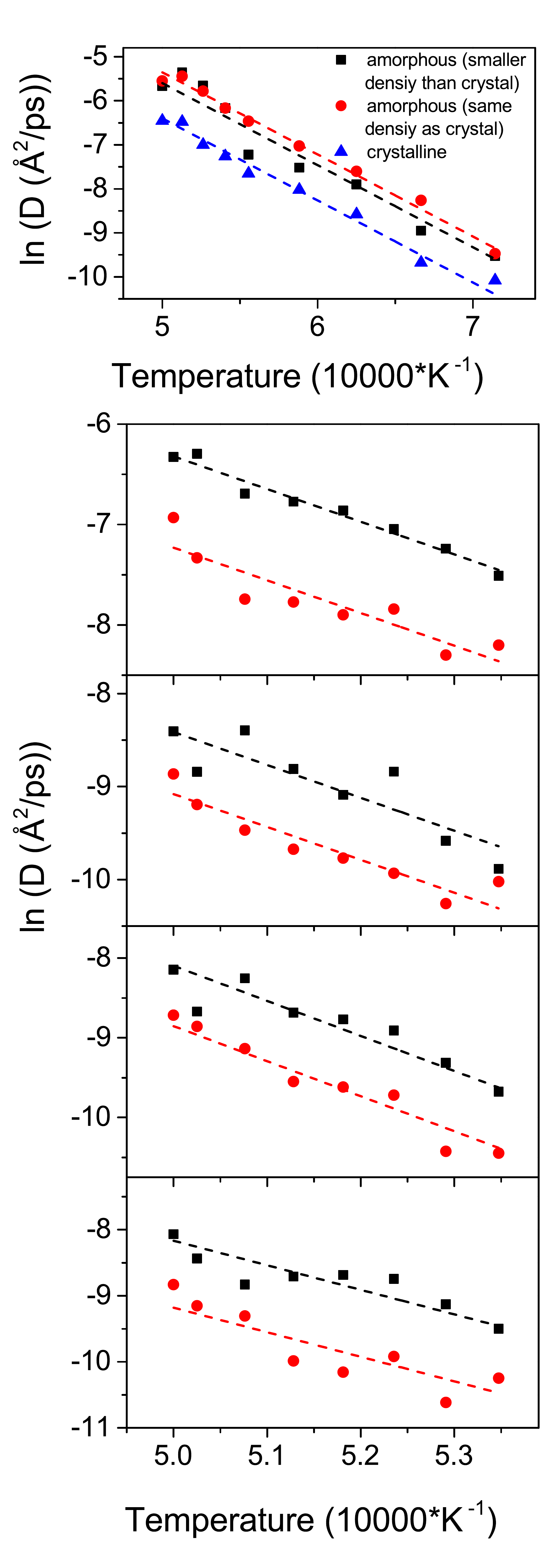}}
\caption{$\ln(D)$ vs inverse temperature, $\frac{1}{T}$ for different atomic species. The range of temperatures in the x-axis is 1400-2000 K for O and 1870-2000 K for cation. The dashed lines are fits of both sets of points to the straight lines assuming that the slopes are the same within the errors present.
}
\label{4}
\end{figure}

\end{document}